\documentclass[%
 aip,
 amsmath,amssymb,
 reprint,%
]{revtex4-1}

\usepackage{graphicx}% Include figure files
\usepackage{dcolumn}% Align table columns on decimal point
\usepackage{bm}% bold math
\usepackage[utf8]{inputenc}
\usepackage[T1]{fontenc}
\usepackage{mathptmx}
\usepackage{etoolbox}
\usepackage{xcolor}

\usepackage{titlesec}

\usepackage{soul}
\usepackage{ulem} % For strikethrough text
\usepackage{cancel}
\newcommand{\newtext}[1]{{#1}} % Highlight new text
\titlespacing*{\section}{0pt}{0.05\baselineskip}{0.05\baselineskip} 
\titlespacing*{\subsection}{0pt}{0.05\baselineskip}{0.05\baselineskip}
\makeatletter
\def\@email#1#2{%
 \endgroup
 \patchcmd{\titleblock@produce}
  {\frontmatter@RRAPformat}
  {\frontmatter@RRAPformat{\produce@RRAP{*#1\href{mailto:#2}{#2}}}\frontmatter@RRAPformat}
  {}{}
}%

\makeatother

\begin{document}

\preprint{AIP/123-QED}

%\title[Exploring active polymer behavior]{Exploring active polymer behavior: Tangential versus Push-Pull models in two dimensions}
\title[Exploring models of active polymer in two dimensions]{Active polymer behavior in two dimensions: A comparative analysis of Tangential and Push-Pull Models}
% Force line breaks with \\
\author{Giulia Janzen}
  \affiliation{ Department of Theoretical Physics, Complutense University of Madrid, 28040 Madrid, Spain}

 \author{Juan Pablo Miranda}
\affiliation{Departamento de Estructura de la Materia, F\'isica T\'ermica y Electr\'onica, Universidad
Complutense de Madrid, 28040 Madrid, Spain}
\affiliation{GISC - Grupo Interdisciplinar de Sistemas Complejos 28040 Madrid, Spain}
\author{J. Martín-Roca}
  \affiliation{Departamento de Estructura de la Materia, F\'isica T\'ermica y Electr\'onica, Universidad
Complutense de Madrid, 28040 Madrid, Spain}
\affiliation{GISC - Grupo Interdisciplinar de Sistemas Complejos 28040 Madrid, Spain}

\author{\newtext{Paolo Malgaretti}}
\affiliation{\newtext{Helmholtz Institute Erlangen-Nurnberg for Renewable Energy (IET-2), Forschungszentrum
J\"ulich, Erlangen, Germany}}
\author{\newtext{Emanuele Locatelli}}
\affiliation{\newtext{Department of Physics and Astronomy, University of Padova, 35131 Padova, Italy}}
\affiliation{\newtext{INFN, Sezione di Padova, via Marzolo 8, I-35131 Padova, Italy}}

\author{Chantal Valeriani}
  \affiliation{Departamento de Estructura de la Materia, F\'isica T\'ermica y Electr\'onica, Universidad
Complutense de Madrid, 28040 Madrid, Spain}
\affiliation{GISC - Grupo Interdisciplinar de Sistemas Complejos 28040 Madrid, Spain}
\email{cvaleriani@ucm.es}
\author{D. A. Matoz Fernandez}
 \affiliation{ Department of Theoretical Physics, Complutense University of Madrid, 28040 Madrid, Spain}
 \email{dmatoz@ucm.es}

\date{\today}% It is always \today, today,
             %  but any date may be explicitly specified

\begin{abstract}

In this work, we compare the structural and dynamic behavior of active filaments in two dimensions using tangential and push-pull models, including a variant with passive end monomers. These models serve as valuable frameworks for understanding self-organization in biological polymers and synthetic materials. At low activity, all models exhibit similar behaviors. Differences emerge in the intermediate range as activity increases, though at higher activity levels, their behaviors converge. Importantly, adjusting for differences in mean active force reveals nearly identical behavior across models. Our results highlight the importance of force definitions in active polymer simulations and provide insights into phase transitions across varying filament configurations.
\end{abstract}

\maketitle

\section{\label{sec:intro}Introduction}
Filamentous and polymeric structures are crucial in both biological and artificial systems. They are found across a range of scales, from microscopic components such as microtubules and actin filaments in the cellular cytoskeleton~\cite{bray2000cell, Schaller2010, ganguly2012cytoplasmic}, to bacteria~\cite{Yaman2019}, worms~\cite{deblais2020phase, deblais2020rheology, nguyen2021emergent}, granular systems~\cite{wen2012polymerlike, soh2019self}, and even macroscopic robotic systems~\cite{marvi2014sidewinding, ozkan2021collective}. These structures exhibit a wide range of dynamic behaviors, including self-organization, motility, and responsiveness to external stimuli, which are fundamental to understanding processes like cell division, tissue development, and artificial soft robotics. In recent years, it has become increasingly apparent that the concept of active matter offers a robust framework for understanding the physics of living systems across these diverse scales. Active matter encompasses self-driven systems capable of converting stored or environmental energy into directed motion, leading to collective phenomena. This includes studies of intracellular components~\cite{Schaller2010, Sumino2012}, artificial self-propelled particles~\cite{Narayan2007, Theurkauff2012, Palacci2013}, microorganisms~\cite{Saw2017, Beer2019}, large animal species~\cite{Sumpter2010, Bialek2012}, robot swarms~\cite{Rubenstein2014, Wang2021}, and filamentous matter~\cite{IseleHolder2015, Winkler2020, philipps2022tangentially, Deblais2023, ubertini2024universal, malgaretti2024coil}.\\
\indent Recently, much attention has been given to the emergent properties of systems combining the activity and conformational degrees of freedom in filamentous and polymeric structures~\cite{schaller2010polar, Loi2011,suma2014motility,Huijun2014A,HuijunB, jiang2014motion, IseleHolder2015,eisenstecken2017internal,duman2018collective,Prathyusha2018, Bianco2108,Shee2021, Prathyusha2022, zhu_non-equilibrium_2024}. Known as active polymers~\cite{winkler2017active}, these systems exhibit dynamics where active forces on individual monomers dictate overall behavior, leading to phenomena such as non-equilibrium phase transitions, dynamic phase separation, and anomalous diffusion \cite{MarchettiRev,romanczuk2012active,PhysRevX.5.031003,thampi2016active,alert2022active,MICHELETTI20111,TUBIANA20241}.\\
\indent Discrete polar active filaments have been modeled using various representations of active forces~\cite{jiang2014motion, IseleHolder2015,Bianco2108}, including tangential forces along the polymer backbone~\cite{Bianco2108} and push-pull forces~\cite{IseleHolder2015}. These models converge to the same continuum limit, making results independent of the discretization method~\cite{Winkler2020}.
However, deviations \newtext{may} emerge when activity surpasses a threshold, especially in systems with weak bond potentials, large bond-length variations, or small numbers of monomers. These deviations underscore the need to examine both microscopic details and system dynamics in highly active regimes. Despite these studies, as far as we are aware a direct comparison between tangential and push-pull models is still lacking, and understanding how different active force definitions influence structure and dynamics is crucial, particularly where deviations from continuum models are expected.\\
\indent In this work, we compare tangential and push-pull models of active filaments in two dimensions, 
\newtext{introducing} a version of the push-pull where the first and last monomers are passive \newtext{to bridge between the two.} 
We analyze these models varying bending rigidity and polymer length in single-filament systems \newtext{as well as at finite density}. As \newtext{it} is well known, \newtext{at infinite dilution} the system undergoes a transition from open chains to spirals as activity increases \cite{IseleHolder2015, Prathyusha2018, gompper20202020}. \newtext{We find that} at low activity, all models exhibit similar structural and dynamic properties. However, as the system approaches the spiral phase, small differences emerge. Nonetheless, in the spiral phase, the behavior converges across all models, regardless of filament length or bending rigidity.\\
\indent At intermediate densities, we explore phase transitions, including the reentrant phase, previously identified in the push-pull model at high activities~\cite{janzen2024density}: this phase transitions from spirals back to open chains. We extend this analysis to the tangential and push-pull models with passive head and tail, revealing that while all models exhibit similar phases, the reentrant phase in the tangential model occurs at higher 
activities due to its smaller active force. Once this force difference is accounted for, the tangential model behaves similarly to the push-pull models at intermediate and high bending rigidities. At low bending rigidity, differences in activity definitions cause slight shifts in the spiral transition. Overall, in two dimensions, once scaled for active force differences, all models behave nearly identically.

\section{Methods}\label{sec:simdet}
\subsection*{Model}
We simulate a system of active filaments, each consisting of $N_b$ beads. We ignore long-range hydrodynamic interactions and consider the dry limit, where damping from the medium dominates, so the surrounding fluid provides only single-particle friction. Thus, the equation of motion for each filament is given by Langevin dynamics, ~\cite{suma2014motility,IseleHolder2015, Prathyusha2018}
\begin{equation}
m_i \, \vec{\ddot{r}}_i=-\gamma \, \vec{\dot{r}}_i+\vec{f}^a_i+\sum_{j\neq i } \vec{f}_{ij}+\vec{R}_i(t),
\end{equation}
where $m_i$ is the mass of bead $i$, $\vec{r}_i=(x_i,y_i)$ represents the bead's spatial coordinates and the dot denotes the time derivatives, $\gamma$ is the damping coefficient and $\langle \vec{R}_i(t) \cdot \vec{R}_j(t^\prime) 
 \rangle= 4 \gamma k_B T \delta_{ij} \delta(t-t^\prime)$ is a delta-correlated random force with zero mean and variance, where  $k_B$ is the Boltzmann constant and $T$ the temperature. Moreover, $\vec{f}_{ij}=-\nabla_i \phi(r_{ij})$ is the interaction force between beads $i$ and $j$, where $r_{ij}=\left\vert\vec{r}_i-\vec{r}_j \right\vert$ and $\phi$ is the bonded and short-range nonbonded pair potentials $\phi=\phi_B+\phi_{NB}$. Bonded interactions $\phi_B=\phi_{bond}+\phi_{bend}$ account for both chain stretching, modeled by the Tether bond potential~\cite{Noguchi2005} and bending modeled with the harmonic angle potential~\cite{Prathyusha2018}. Next, the nonbonded interactions, $\phi_{NB}$ account for steric repulsion and are modeled with the Weeks-Chandler-Anderson (WCA) potential~\cite{weeks1971role}. All details regarding the bonded and nonbonded interaction parameters are described in the Supplementary Material~\cite{SI}. Lastly, $\vec{f}^a_i$ represents the active self-propulsion force on bead $i$. 
 
In this study, we focus on employing two models of active polymers: the tangential model and the push-pull model. While both models incorporate tangential propulsion, they differ fundamentally in the way active forces are applied along the polymer backbone. The tangential model introduces activity uniformly along the backbone of the polymer, with the active force aligned with the local tangent of the filament. This design reflects biological scenarios where activity is distributed consistently along the filament’s contour. Importantly, in this model, the extremities of the polymer are passive, meaning no active force acts on the first and last monomers. In contrast, the push-pull model applies active forces at each monomer as a combination of contributions from the adjacent bonds. This results in a “push-pull” effect, where activity arises from the interplay of forces on either side of a monomer along the polymer contour. At the extremities, the absence of a neighboring bond naturally reduces the activity at the polymer’s head and tail. This feature makes the push-pull model particularly suited to capture systems where forces originate locally from bond-level interactions.

\begin{figure}[h!]
\includegraphics[width=\columnwidth]{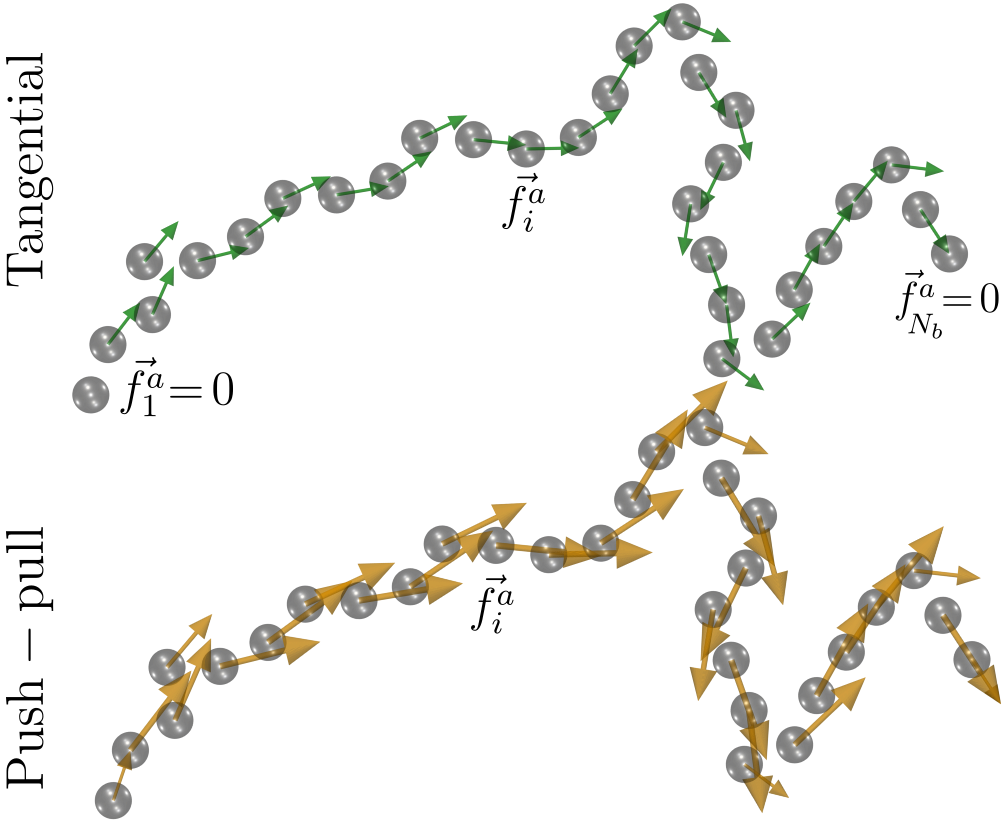}
\caption{Schematic representation of the self-propelled semiflexible filament model. For simplicity, bonded and nonbonded interactions are omitted. Grey dots represent the monomers, while arrows indicate the active force $\vec{f_i^a}$. In the tangential model, the active force acts along the polymer backbone, tangential to the filament. The active force applied to each monomer is uniform (hence the identical arrows), except for the first and last monomers, which are passive. In the push-pull model, the active force is directed along the tangent to the filament contour at each monomer $i$. The active force varies between monomers, with larger arrows indicating stronger activity and smaller arrows representing weaker activity. To avoid redundancy, the push-pull model with passive head and tail is not shown; it is the same as the push-pull model, except the filament's head and tail are passive.}
\label{fig:scheme}
\end{figure}

\noindent
\textbf{The Tangential model:} 
The first model we consider is the tangential (T) model, proposed by Bianco \textit{et al.} \cite{Bianco2108}. In this model, polymer activity is introduced by a tangential self-propulsion force applied along the polymer backbone (see Fig.~\ref{fig:scheme}
(a)). The active force on each monomer $i$ is defined as:
\begin{equation}
    \vec{f_i^a} = f_a \, \frac{{\vec r}_{i+1} - {\vec r}_{i-1}}{|{\vec r}_{i+1} - {\vec r}_{i-1}|} = f_a \, \vec{t}_{i-1,i+1},
\end{equation}
with $f_a$ the magnitude of the active force directed along the vector ${\vec r}_{i+1} - {\vec r}_{i-1}$, which is parallel to the tangent of the polymer backbone \cite{Bianco2108,foglino2019}, and $\vec{t}_{i-1,i+1}$ the unit tangent vector. In this model, the first and last beads of the polymer are passive, meaning that no active force acts on them.

\noindent
\textbf{Push-pull model:} 
The second model we examine is the well-known push-pull (PP) model~\cite{IseleHolder2015,Prathyusha2018} (see Fig.~\ref{fig:scheme}(b)), where the active force $\vec{f}^a_i$ on bead $i$ is given by
\begin{equation}
   \vec{f}^a_i= f_a (\vec{t}_{i-1,i}+\vec{t}_{i,i+1}), 
\end{equation}
as before, here $f_a$ represents the magnitude of the active force, and $\vec{t}_{i,i+1} = \vec{r}_{i,i+1} / r_{i,i+1}$ is the unit tangent vector along the bond connecting beads $i$ and $i+1$. Note that, in contrast to the T-model, the first and last beads experience reduced activity compared to the others, as they only receive contributions from a single neighboring bond.

\noindent
\textbf{Push-pull with passive head and tail:} 
Finally, we introduce the push-pull model with passive head and tail (PHT), which is similar to the standard push-pull model, except the first and last beads are passive. Since the T-model also features passive end beads, this variant allows \newtext{to bridge} between the tangential and push-pull models.

\subsection*{Simulation Details}
In this study, we focus on polymers with a degree of polymerization \(N_b = 50\), with a monomer size \(\sigma = 1.0\), interaction energy scale \(\epsilon = 1.0\), thermal energy \(k_B T = 10^{-1}\), \newtext{mass $m=1.0$}, and a damping coefficient \(\gamma = 1.0\)\newtext{, i.e, underdamped limit}~\cite{janzen2024density}. The length of each active polymer is calculated as \(L \approx b \,(N_b - 1)\sigma\), where \(b = 0.86\). Note that all parameters are given in LJ units. We define two key dimensionless numbers for our analysis: the scaled persistence length \(\xi_p/L = 2b\kappa/(Lk_B T)\), representing the filament's thermal (passive) stiffness, and the active Péclet number \(Pe = f_a L^2 / (\sigma k_B T)\), which characterizes the balance between active forces and thermal fluctuations. The bending rigidity \(\kappa\) is held constant by setting \(\xi_p/L = 0.06, 0.3, 1.4\). To investigate filaments in suspension, we set the packing fraction to \(\rho = N_f N_b \,\pi\,\sigma^2/4L_{\text{box}}^{2} = 0.4\), with the number of filaments \(N_f = 10^{3}\), and varied the Péclet number \(Pe\). Additionally, we explored the single-filament case with polymerization degrees \(N_b = 5\) and \(N_b = 15\); the results for these cases, along with further details, are provided in the Supplementary Material~\cite{SI}.

Simulations were performed using the GPU-accelerated SAMOS molecular dynamics package~\cite{SAMoS2024}, employing the BAOAB Langevin scheme for numerical integration~\cite{leimkuhler2015molecular}. Data analysis was carried out with custom Python scripts, and visualizations were generated using ParaView~\cite{fabian2011paraview}. 

\section{\label{sec:results} Results and discussion}

\subsection*{Single filament}

We begin comparing the three models by examining the structural properties of the systems in the single-filament case. The fist structural property we compute to characterize the behavior of the three models as a function of the Péclet number and to compare them, we compute the turning number~\cite{krantz1999handbook,Shee2021}, denoted as $\psi$.
 This quantity is defined as:

\begin{equation}
  \psi = \frac{1}{2 \pi} \left\vert \sum_{j=1}^{N_b-1} (\theta_{j+1}-\theta_j)\right \vert,
\label{eq:turning_number}
\end{equation}
where $(\theta_{j+1} - \theta_j)$ represents the angular variation between two consecutive monomers, and $N_b$ is the total number of monomers in the polymer. The turning number $\psi$ measures the number of turns the chain makes between its two ends. For a straight chain, $\psi = 0$; for a circular configuration, $\psi = 1$;  higher values of $\psi$ correspond to the presence of loops or spirals.

{Furthermore, because the active force in the push-pull models (PP and PHT) is defined differently from that in the tangential model, we compute the curvature to assess how these variations in force definitions influence filament conformation. The curvature $K$ is defined as:
\begin{equation}
    K = \frac{x' y'' - y' x''}{(x'^2 + y'^2)^{\frac{3}{2}}},
\end{equation}
where $x$ and $y$ represent the coordinates of the beads, and the primes denote first and second derivatives, respectively. 
All these properties are averaged over $10^2$ distinct decorrelated snapshots at steady state and over all the filaments $N_f = 10^3$.

Figure~\ref{fig:struc_single_fil} shows the turning number and curvature for a single filament with \(N_b = 50\) at three different bending rigidities: \(\xi_p/L = 0.06, 0.3,\) and \(1.4\).
\begin{figure}[h!]
\includegraphics[width=8.5cm]{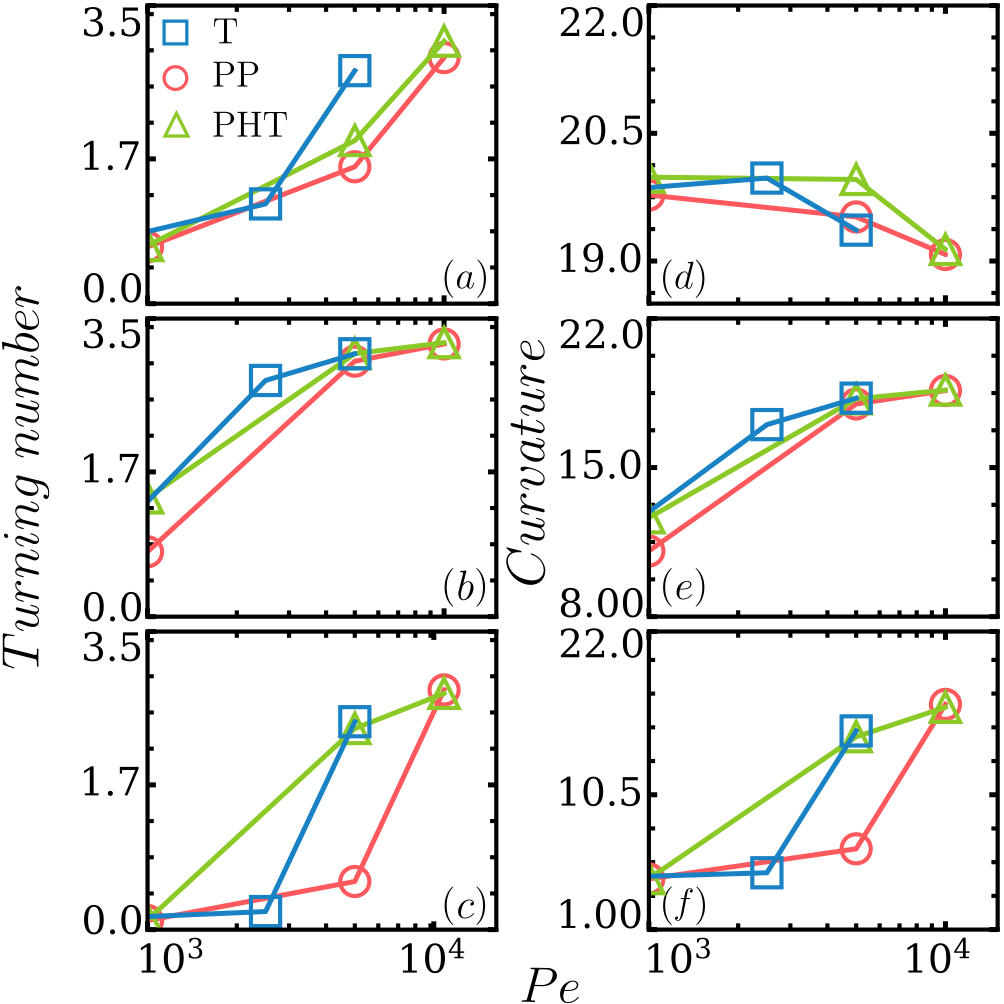}
\caption{Structural properties of the single filaments with $N_b=50$, averaged over the number of filaments $N_f$, as a function of the P\'eclet number $Pe$ for the three models: blue squares represent the T model, red circles the PP model, and green triangles  the PHT model.
(a), (b) and (c)  panels represent the turning number, $\langle \vert\psi\vert \rangle$, for $\xi_p/L=0.06,0.3,$ and $1.4$, respectively.
(d), (e) and (f) panels represent the  curvature, $\langle \sum |K| \rangle$, for $\xi_p/L=0.06,0.3,$ and $1.4$, respectively. Note that the minimum value on the y-axis varies across the three panels.
The three models exhibit the same behavior in both turning number and curvature: initially forming open chains,  as $Pe$ increases spirals begin to form. The only distinction between the models lies in the specific $Pe$ values at which the filament enters a stable spiral phase. However, once this stable spiral phase is reached, the turning number and curvature are identical across all three models.
}
\label{fig:struc_single_fil}
\end{figure}
For both structural properties, the three models exhibit the same qualitative behavior. At low \(Pe\), the polymers form open chains, and as activity increases, the filaments transition into spirals (see Supplementary Material~\cite{SI}). Since higher turning number and curvature values correspond to more compact structures, we observe that before the spiral transition, the PHT model shows \newtext{a} slightly higher degree of compactness compared to the other two models, while the T model has the least compact structures. However, once the filaments enter the spiral phase (\(\psi > 3\)), all three models behave similarly. Finally, we observed similar behavior for shorter \newtext{and longer} polymer lengths (see Supplementary Material~\cite{SI}).

To properly compare the three models, it is important to study not only their structural properties but also their dynamics. Thus, we computed the mean-squared displacement, denoted as MSD$(t)$, after ensuring the system is equilibrated,
\begin{equation}
\mathrm{MSD}(t) = \left \langle  |\vec{r}_{cm}(t)-\vec{r}_{cm}(t_0)|^2  \right \rangle,
\end{equation}
where  $\vec{r}_{cm}(t)$ represents the position of the filament's center of mass at time $t$,  $\vec{r}_{cm}(t_0)$ is its position at the initial time $t_0$, and the $\left \langle \dots \right \rangle$ represent the ensemble averages over $10^3$ different filament configurations. This dynamical property quantifies the displacement of the filament's center of mass in real space relative to its initial position. By analyzing the MSD, we can determine whether the system exhibits diffusive behavior or anomalous dynamics.

\begin{figure}[h!]
\includegraphics[width=8.5cm]{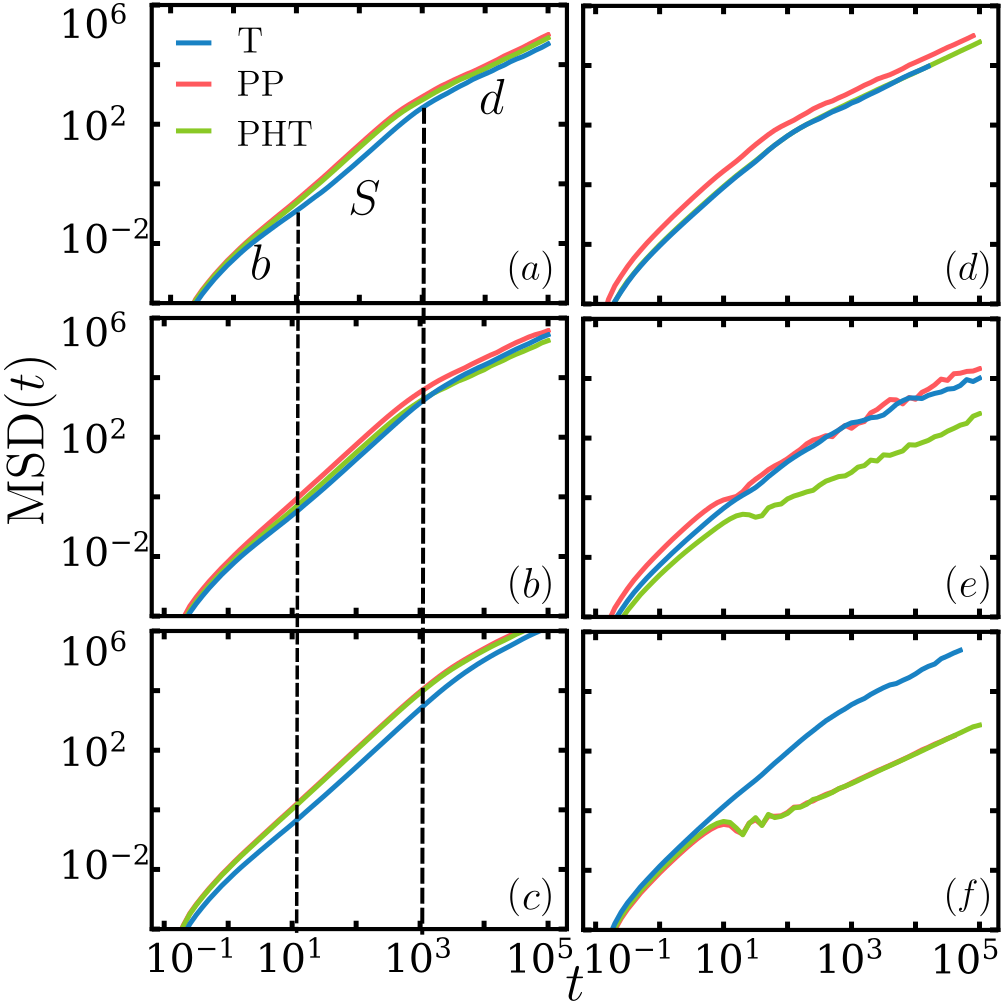}
\caption{Mean squared displacement MSD(t) as a function of time for the three models: blue line represent the T model, red line for the PP model, and green line for the PHT model.
(a), (b) and (c)  MSD(t) at $Pe=10^3$, for $\xi_p/L=0.06$, $\xi_p/L=0.3$ and $\xi_p/L=1.4$, respectively. The mean-squared displacement of the three models are identical, regardless of the bending rigidity.We identify and mark the three regimes observed in the MSD plot as follows: the ballistic regime  (denoted as \textit{b} in the plot) for $t<10^1$; superdiffusive (denoted as \textit{S} in the plot) for $10^1<t<10^3$ characterized by a slope of approximately 1.5 for $\xi_p/L=0.06, 0.3$, and 1.8 for $\xi_p/L=1.4$; and finally, diffusive (denoted as \textit{d} in the plot) for $t>10^3$.
(d), (e) and (f) MSD(t) at $Pe=10^4$, for $\xi_p/L=0.06$, $\xi_p/L=0.3$ and $\xi_p/L=1.4$, respectively. For $Pe=10^4$, we identify different behaviors depending on the bending. (d) For $\xi_p/L=0.06$, similarly to the smaller P\'eclet ($Pe=10^3$), we observe a ballistic regime for $t<10^0$, followed by a superdiffusive behavior for $10^0<t<10^2$ with a slope of approximately 1.5 followed by a diffusive regime for  $t>10^2$. (e) For $\xi_p/L=0.3$, we observe a ballistic regime for $t<10^0$, followed by a superdiffusive behavior for $10^0<t<10^2$ with a slope of approximately 1.4 for the T and PP model. In contrast, we observe a subdiffusive behavior with a slope of 0.8 for the PHT model, followed by a diffusive regime for  $t>10^2$. (f) For $\xi_p/L=1.4$, we observe a ballistic regime for $t<10^0$, followed by a superdiffusive behavior for $10^0<t<10^2$ with a slope of approximately 1.8  for the T model. In contrast, we observe a subdiffusive behavior with a slope of 0.5 for the PP and PHT model, followed by a diffusive regime for  $t>10^2$. }
\label{fig:msd}
\end{figure}

Figures~\ref{fig:msd}(a), (b), and (c) display the MSD at \(Pe = 10^3\) for \(\xi_p / L = 0.06, 0.3, 1.4\), respectively. In this case, the three models exhibit the same dynamical behavior, independent of the bending rigidity. This consistency aligns with the structural properties, which at this \(Pe\) are identical for small and large bending rigidities, with only slight differences at intermediate bending rigidity.

In the spiral phase, at \(Pe = 10^4\), as shown in Fig.~\ref{fig:msd}(d), (e), for \(\xi_p / L = 0.06, 0.3\), the spirals in the PHT model exhibit slower dynamics, although the MSD slope remains the same for all three models. The T and PP models, on the other hand, have nearly identical dynamics. At higher bending rigidity, \(\xi_p / L = 1.4\), the filaments in the T model (Fig.~\ref{fig:msd}(f)) exhibit faster dynamics and a different MSD slope compared to the PP and PHT models, which share similar dynamics. This difference arises because, at this particular $Pe$, the spiral phase in the T model is not yet stable, and a larger fraction of the filaments remain in the open-chain state, which is characterized by faster dynamics. As a result, the MSD slope in the T model differs from the PP and PHT models, where most filaments are in the spiral phase. The lower turning number for the T model (Fig.~\ref{fig:struc_single_fil}(c)) further supports this, indicating a higher proportion of open chains. Since the MSD is averaged over multiple filaments, the presence of these fast-moving open chains in the T model leads to overall faster dynamics and distinct MSD behavior compared to the other two models.
Overall, in the single-filament case, the three models display very similar behavior, with only a minor shift in the spiral transition for the T model.

\subsection*{Filament in suspension}

To further investigate the differences between the three models, we now examine filaments in suspension at a packing fraction of \(\rho = 0.4\).
Figure~\ref{fig:struc_melt} presents the behavior of the averaged properties, specifically the turning number \(|\psi|\) and the curvature \(\langle \sum |K| \rangle\), as a function of the Péclet number for all three models. The figure focuses on results for \(Pe \geq 10^3\), as the models display similar behavior across all bending rigidity values at lower Péclet numbers (i.e., \(Pe < 10^3\)).
\begin{figure}[h!]
\includegraphics[width=8.5cm]{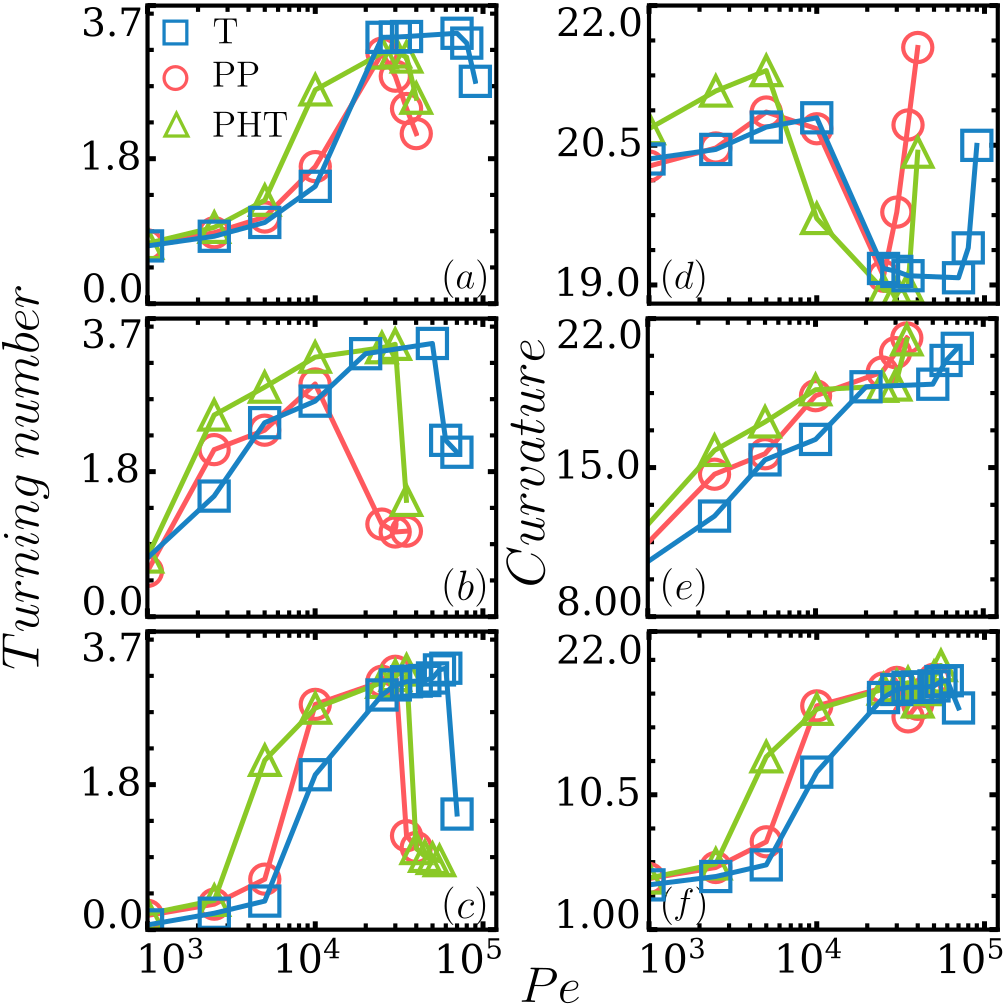}
\caption{Structural properties of the filaments, averaged over the number of filaments $N_f$, as a function of the P\'eclet number $Pe$ for the three models: blue squares represent the T model, red circles for the PP model, and green triangles for the PHT model.
(a), (b) and (c)  The turning number, $\langle \vert\psi\vert \rangle$, for $\xi_p/L=0.06, 0.3$ and $1.4$, respectively.
(d), (e) and (f) The Curvature, $\langle \sum |K| \rangle$, for $\xi_p/L=0.06, 0.3$ and $1.4$, respectively. Note that the minimum value on the y-axis varies across the three panels. The three models display similar behavior in both turning number and curvature: they initially form open chains, and as $Pe$ increases, all filaments transition into spirals. With further increases in $Pe$, the spirals begin to open up, leading to a reentrant phase. (a), (d) For small bending rigidity ($\xi_p/L=0.06$), the T and PP models behave similarly, while the HTP model reaches the spiral phase slightly earlier than the other two. (b), (c) As bending rigidity increases ($\xi_p/L=0.3$, and $1.4$), the PP model is the first to exhibit the reentrant phase, whereas, in the T and HTP models, this phase occurs at higher $Pe$ values.}
\label{fig:struc_melt}
\end{figure}

Figures~\ref{fig:struc_melt}(a) and (d) show the turning number and curvature, respectively, for a bending rigidity of \(\xi_p/L = 0.06\). At this bending rigidity, all three models exhibit very similar behavior. However, the push-pull model with a passive head and tail displays slight deviations from the other two models around \(Pe = 10^4\), due to a higher number of spirals in the system compared to the PP and T models. When the Péclet number increases to approximately \(Pe = 2.5 \times 10^4\), all filaments form spirals across all three models. For even higher \(Pe\) values (\(Pe > 2.5 \times 10^4\)), the two push-pull models exhibit the same behavior: the turning number begins to decrease as some spirals start to open, indicating the onset of a reentrant phase\cite{janzen2024density}, \newtext{due to inertia}. The reentrant phase occurs \newtext{also in the T model, but} at even higher Péclet numbers. \newtext{In contrast}, as shown in Figure \ref{fig:struc_melt}(d), the curvature changes only slightly with increasing Péclet number. This minimal variation occurs because the small bending allows the filaments to \newtext{attain very compact and bent conformations}, even at low Péclet numbers.

Figures~\ref{fig:struc_melt}(b) and (e) represent the turning number and curvature for $\xi_p/L=0.3$, respectively. The results show that, for all three models, the turning number increases with increasing P\'eclet number and then decreases again at very high $Pe$ (i.e., $Pe > 10^4$). At $Pe \approx 10^4$, all filaments form spirals in the system across all models, though the spirals in the THP model are slightly more compact than in the other two models.
All three models exhibit a non-monotonic behavior in $\psi$, characteristic of \newtext{the same} reentrant phase \cite{janzen2024density} \newtext{as before}. However, the reentrant phase occurs \newtext{now} at different $Pe$ values for each model. In the T and PHT models, spirals open up at larger $Pe$ compared to the PP model, where the spirals begin to open at $Pe \approx 2.5 \times 10^4$. Although both the T and PHT models experience a shift in the reentrant phase relative to the PP model, the phase occurs at even higher $Pe$ values in the \newtext{T} model than in the \newtext{PHT} model. As discussed in the previous section (Fig.~\ref{fig:msd}(e)), the dynamics in the spiral phase for \newtext{isolated chains in} the PHT model is slower than in the other two models, leading to less frequent movement and collisions of the spirals. This slower dynamic may explain why a higher $Pe$ is needed, in the PHT model \newtext{at intermediate values of the bending rigidity,} to open the spirals and observe a reentrant phase.

Additionally, the curvature (Figure~\ref{fig:struc_melt}(e) ) in the T model is slightly lower than in the other two models, which could be attributed to the different definition of the active force. This variation affects the angle distribution between three consecutive monomers. Overall, the curvature behaves similarly across all three models as a function of the P\'eclet number: as $Pe$ increases, the curvature also increases, indicating that the filaments become more compact. The curvature is less sensitive in this context, whereas the turning number clearly emphasizes the reentrant phase.

Finally, Fig.~\ref{fig:struc_melt} (c) and (f) represent the turning number and curvature for $\xi_p/L=1.4$, respectively.
Both panels show that, although all three models exhibit the same qualitative behavior in both turning number and curvature, the pure spiral phase (where all filaments in the system form spirals) and the reentrant phase in the T model are shifted by approximately a factor of two compared to the PP and PHT models. It is important to notice, that this factor depends on the bending rigidity as it is clearly shown in Fig.~\ref{fig:struc_melt}(d), and thus, this is not constant.

In summary, these results indicate that the three models share the same qualitative behavior: starting with an open chains phase, followed by a pure spiral phase, and finally a reentrant phase. Consistent with findings at higher density by Prathyusha \textit{et al.}~\cite{Prathyusha2018}, in all three models the spiral transition occurs at higher $Pe$ for both small and large bending rigidities ($\xi_p/L=0.06, 1.4$) compared to intermediate bending rigidity ($\xi_p/L=0.3$). The only difference between the models is quantitative, as the $Pe$ values at which the spiral and reentrant phases occur are shifted.

\subsection*{Understanding Model Differences}

Since the active force in the tangential model is defined differently from the push-pull models (PP and PHT), we compute the mean force \newtext{over} each bead to better understand the differences between these models. Specifically, the spiral transition arises from a competition between the active and bending forces ~\cite{Prathyusha2018}. Building on this understanding, we focus on the combined contribution of these forces, averaging them across all filaments in the system. Furthermore, to highlight the differences in the forces acting on filaments under the PP model compared to the T and PHT models, we normalize this combined mean force using the values obtained in the PP model. \cite{Prathyusha2018}.
\newtext{Figures}~\ref{fig:force} (a), (b), and (c) show the mean of the combined active and bending forces for the T and PHT models, normalized by the mean force in the PP model. This is plotted as a function of \(Pe\) for \(\xi_p/L = 0.06, 0.3,\) and \(1.4\), plotted in the top, middle and bottom panel respectively. 
\begin{figure}[h!]
\includegraphics[width=8.5cm]{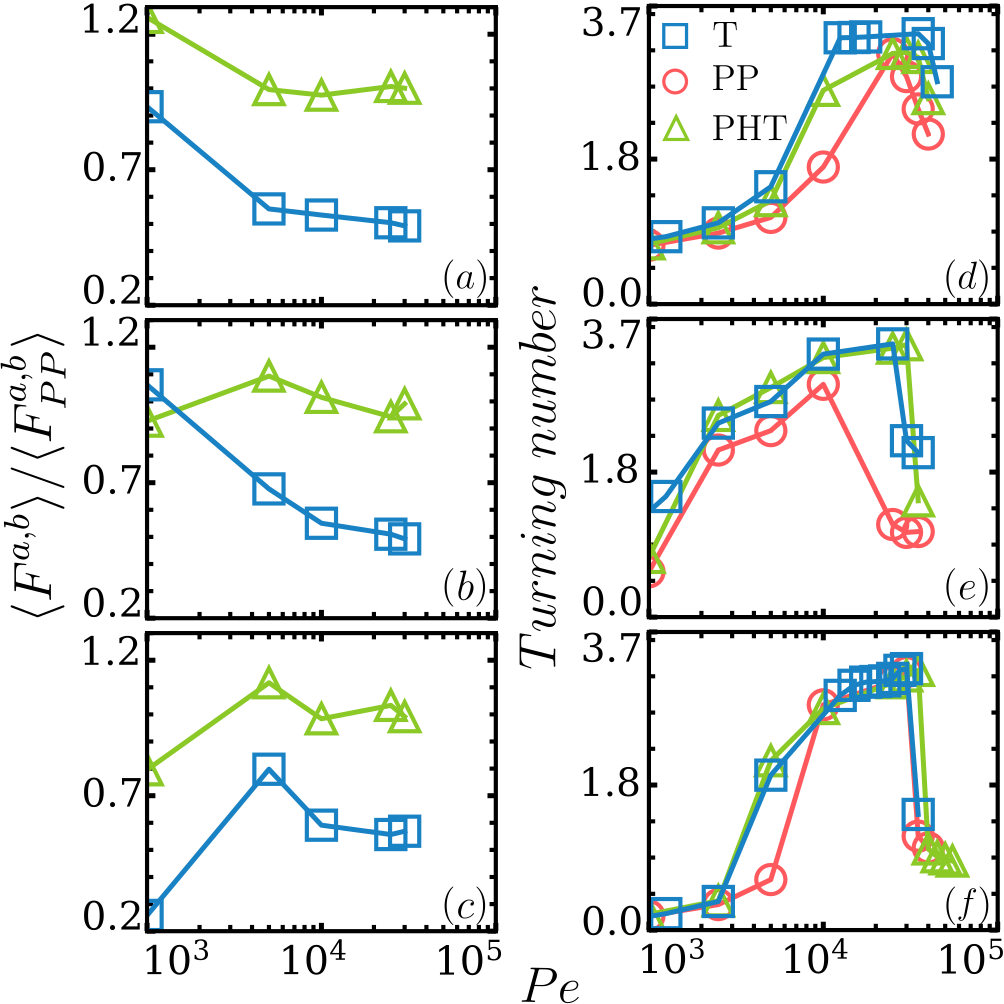}
\caption{(a), (b), and (c) The mean of the sum of the active force and the bending force, $\langle F^{a,b} \rangle$, for the T (blue squares) and PHT (green triangles) models normalized by the mean active force of the PP model, $\langle F^{a,b}_{PP} \rangle$, for $\xi_p/L = 0.06$, 0.3, and 1.4, from top to bottom respectively. This ratio exhibits similar behavior, though shifted, for both the PHT and T models across all bending rigidities. Before the spiral phase ($Pe \approx 10^3$), the bending force contributes to $F^{a,b}$, while in the spiral phase, the active force becomes the dominant contribution. In the PHT model, the active force is nearly identical to that of the PP model, whereas in the T model, it is approximately half of the PP model's active force.
(d), (e), and (f) The turning number as a function of the P\'eclet number $Pe$ for $\xi_p/L=0.06,0.3,$ and $1.4$, from top to bottom respectively. For the T model (blue squares), the activity is approximately half that of the PP (red circles) and PHT (green triangles) models. Consequently, in these simulations, the P\'eclet number used for the T model is twice that of the PP model. For all bending rigidities, when the Péclet number in the T model is twice compared to that used in the PP models, the behavior of the T model closely aligns with that of the PP and PHT models.}
\label{fig:force}
\end{figure}

These figures show that the ratio \(\langle F^{a,b}\rangle/\langle F^{a,b}_{PP}\rangle\) exhibits similar behavior, though shifted, for both the PHT and T models across all bending rigidities. The shift occurs because, in the PHT model, the active force is almost identical to that of the PP model, while in the T model, the active force is consistently about half that of the PP model under the parameters used in our simulations (see Supplementary Material \cite{SI}). Before the spiral transition, both the active and bending forces contribute to \(F^{a,b}\), but in the spiral phase, the bending contribution becomes minimal. This explains why, for \(Pe > 10^4\), the ratio \(\langle F^{a,b}\rangle/\langle F^{a,b}_{PP}\rangle\) approaches 1 in the PHT model and approximately 0.5 in the T model. This difference in the mean force accounts for the shift in the \(Pe\) values at which the spiral transition and reentrant phase occur compared to the PP model.

Figure~\ref{fig:force} (d), (e), and (f) show results similar to Fig.~\ref{fig:struc_melt}, but with the \(Pe\) values for the T model rescaled to be twice those used for the PP and PHT models. These figures demonstrate that, at intermediate and high bending rigidities (\(\xi_p = 0.3, 1.4\)), when accounting for the higher active force in the PP and PHT models, the T model behaves identically to the PP model.
However, at low bending rigidity (\(\xi_p = 0.06\)), the transitions in the T model are slightly shifted compared to the PP and PHT models. In this low-bending regime, the filament's increased flexibility makes the differences in the active force definition more pronounced. As the bending rigidity increases, both the tangential and push-pull models exhibit active forces tangential to the polymer, minimizing the impact of force definition differences on the structure and dynamics of the systems.

In conclusion, the active force in the T model is consistently smaller than that in the PP model by a constant factor. When this factor is considered, the models exhibit the same behavior at intermediate and large bending rigidity values. At small bending, the models still share the same qualitative behavior, though the spiral transition and reentrant phase are slightly shifted.

\section{\label{sec:conclusions}Conclusions}

In conclusion, our study highlights that the primary difference between the tangential model and the push-pull models is due to the smaller mean active force in the tangential model. Through comprehensive simulations of single filaments with varying lengths and polymer melts across a range of bending rigidities, we have shown that this force discrepancy is critical in explaining the observed differences. However, once this difference is accounted for, the tangential model exhibits the same structural and dynamic behavior as the push-pull models, particularly at intermediate and high bending rigidities. These results emphasize that, despite the difference in force definitions, the underlying physics in two dimensions of active filament behavior remains consistent across models when properly scaled.

At infinite dilution and small bending rigidity, the qualitative behavior is similar across all three models, with the only difference being a slight shift of the spiral transition in the T model. We have shown that, either at high bending rigidity or in the spiral phase, the different definitions of the active force lead to similar structural and dynamical behaviors. In contrast, at low bending rigidity, where the filaments are more flexible, the different definitions of activity have a stronger influence on the angle distribution between consecutive monomers, leading to shifts in the spiral \newtext{phase}.
Overall, we conclude that in two dimensions, when the mean active force is the same and the bending rigidity is sufficiently large, the tangential and push-pull models behave essentially identically. However, in three dimensions, we expect the tangential and push-pull models to exhibit different behaviors. In the three-dimensional case, increasing activity in the tangential model tends to lead to a collapsed state \cite{Bianco2108,LocatelliPRL2021}, with respect to the push-pull model. As a result, the different definitions of activity in three dimensions are likely to produce distinct behaviors between the models.

\section{Supplementary materials}

This section includes additional technical details that complement the findings discussed in the main manuscript. Specifically, it provides an in-depth explanation of the polymer models used, including the interaction potentials governing the behavior of bonded and nonbonded interactions in the polymer chains. These include both attractive and repulsive forces, as well as bending stiffness parameters. The document also includes further structural analysis, such as the calculation of the end-to-end distance and the gyration radius for different polymer configurations and models. In addition, supplementary figures illustrate the structural properties, such as the gyration radius and end-to-end distances, across various conditions, expanding on the results presented in the main text.
\section*{Author Contributions}
GJ and JPM contributed equally. GJ, JPM, JMR, CV, and DAMF designed the research. GJ and JPM conducted the research, establishing the methodology, performing the analysis, and validating the data. DAMF developed the software. EL and PM provided critical insights into the role of active forces in the spiral state, the behavior of long filaments, and its implications, as well as contributing to the overall revision of the manuscript. CV and DAMF supervised the project. All authors contributed to writing and editing the manuscript, participated in discussions, and provided meaningful insights.

\section*{Conflicts of interest}
There are no conflicts to declare.

\section*{Acknowledgements}
D.M.F. and G.J. gratefully acknowledge support from the Comunidad de Madrid and the Complutense University of Madrid (Spain) through the Atracción de Talento program, grant 2022-T1/TIC-24007. C.V. acknowledges funding from MINECO under grants IHRC22/00002 and PID2022-140407NB-C21. J.M.R. acknowledges financial support from the UCM
predoctoral contract (call CT15/23).
\section*{Data Availability Statement}
Data available on request from the authors.

\bibliography{biblo} 
\bibliographystyle{unsrt}
\end{document}